\documentclass[a4paper,11pt]{article}
\usepackage{pos}
\usepackage{lineno} 

\title{Lepton Flavour Universality tests in electroweak penguin decays at LHCb}
\ShortTitle{LU tests in EW penguin decays at LHCb}

\author*[a,1]{Carla Marin Benito}

\affiliation[a]{Universit\'e Paris-Saclay, CNRS/IN2P3, IJCLab,\\
   91405 Orsay, France }

\note{On behalf of the LHCb collaboration}

\emailAdd{carla.marin.benito@cern.ch}

\abstract{The coupling of the electroweak gauge bosons of the Standard Model (SM) to leptons is flavour universal. Extensions of the SM do not necessarily have this property. Rare decays of heavy flavour are suppressed in the SM and new particles may give sizeable contributions to these processes, therefore, their precise study allows for sensitive tests of lepton flavour universality. Of particular interest are rare \bsll decays that are well accessible at the LHCb experiment. Recent results from LHCb on lepton flavour universality in rare \bsll decays are discussed.}

\FullConference{%
  40th International Conference on High Energy physics - ICHEP2020\\
  July 28 - August 6, 2020\\
  Prague, Czech Republic (virtual meeting)
}

\usepackage{xspace} 
\usepackage{upgreek} 

\def\PB      {\ensuremath{B}\xspace}                 
\def\PD      {\ensuremath{D}\xspace}                 
\def\PJ      {\ensuremath{{J}}\xspace}          
\def\PK      {\ensuremath{{K}}\xspace}          
       
\def\Pb      {\ensuremath{{b}}\xspace}
\def\Pc      {\ensuremath{{c}}\xspace}                 
                 
\def\Pe      {\ensuremath{{e}}\xspace}

\def\Pp      {\ensuremath{{p}}\xspace}
\def\Ps      {\ensuremath{{s}}\xspace}

\def\Pmu         {\ensuremath{\upmu}\xspace} 
                 
\def\Ppi         {\ensuremath{\pi}\xspace}                 
\def\Ppsi        {\ensuremath{\psi}\xspace}

\mathchardef\PXi="7104
\mathchardef\PLambda="7103
\mathchardef\POmega="710A

\def\squark    {{\ensuremath{\Ps}}\xspace}
\def\bquark    {{\ensuremath{\Pb}}\xspace}
\def\cquark    {{\ensuremath{\Pc}}\xspace}

\def\pion   {{\ensuremath{\Ppi}}\xspace}
\def\piz    {{\ensuremath{\pion^0}}\xspace}

\def\kaon   {{\ensuremath{\PK}}\xspace}

\def\Kstar  {{\ensuremath{\kaon^{*}}}\xspace}
\def\Kstarz {{\ensuremath{\kaon^{*0}}}\xspace}

\def\Kp       {{\ensuremath{\kaon^+}}\xspace}
\def\Km      {{\ensuremath{\kaon^-}}\xspace}

\def\D       {{\ensuremath{\PD}}\xspace}
\def\Dz      {{\ensuremath{\D^0}}\xspace}

\def\B       {{\ensuremath{\PB}}\xspace}
\def\Bd      {{\ensuremath{\B^0}}\xspace}
\def\Bs      {{\ensuremath{\B^0_\squark}}\xspace}
\def\Bu      {{\ensuremath{\B^+}}\xspace}

\def\proton      {{\ensuremath{\Pp}}\xspace}

\def\Lz          {{\ensuremath{\PLambda}}\xspace}

\def\Lb      {{\ensuremath{\Lz^0_\bquark}}\xspace}

\def\pion   {{\ensuremath{\Ppi}}\xspace}
\def\pip    {{\ensuremath{\pion^+}}\xspace}

\def\epem       {{\ensuremath{\Pe^+\Pe^-}}\xspace}

\def\mup        {{\ensuremath{\Pmu^+}}\xspace}
\def\mun        {{\ensuremath{\Pmu^-}}\xspace} 

\def\ellm       {{\ensuremath{\ell^-}}\xspace}
\def\ellp       {{\ensuremath{\ell^+}}\xspace}

\def\jpsi     {{\ensuremath{{\PJ\mskip -3mu/\mskip -2mu\Ppsi\mskip 2mu}}}\xspace}
\def\psitwos  {{\ensuremath{\Ppsi{(2S)}}}\xspace}

\newcommand{\decay}[2]{\ensuremath{#1\!\to #2}\xspace}         

\def\bsll         {\decay{\bquark}{\squark\ellp\ellm}}

\def\btosmm {\decay{\bquark}{\squark\mup\mun}}

\def\BdToKstmm   {\decay{\Bd}{\Kstarz\mup\mun}}

\def\BuToKpmm    {\decay{\Bu}{\Kp\mup\mun}}

\def\LbTopKmm    {\decay{\Lb}{\proton\Km\mup\mun}}

\def\BuToKpee      {\decay{\Bu}{\Kp\epem}}
\def\BdToKstee   {\decay{\Bd}{\Kstarz\epem}}
\def\LbTopKee    {\decay{\Lb}{\proton\Km\epem}}

\def\BuToKll   {\decay{\Bu}{\Kp\ellp\ellm}}

\def\BuToKJpsiee   {\decay{\Bu}{\Kp\jpsi(\to\epem)}}

\def\BdToKstll   {\decay{\Bd}{\Kstarz\ellp\ellm}}
\def\LbTopKll    {\decay{\Lb}{\proton\Km\ellp\ellm}}
\def\BsToPhill  {\decay{\Bs}{\phi\ellp\ellm}}
\def\BuToPill   {\decay{\Bu}{\pip\ellp\ellm}}

\def \RK    {\ensuremath{R_K}\xspace}
\def \RKst {\ensuremath{R_\Kstarz}\xspace}
\def \RpK {\ensuremath{R_{pK}}\xspace}

\def \invrJpsi {\ensuremath{r_\jpsi^{-1}}\xspace}

\newcommand{\tev}{\ensuremath{\mathrm{\,Te\kern -0.1em V}}\xspace}
\newcommand{\mev}{\ensuremath{\mathrm{\,Me\kern -0.1em V}}\xspace}
\newcommand{\mevc}{\ensuremath{{\mathrm{\,Me\kern -0.1em V\!/}c}}\xspace}
\newcommand{\mevcc}{\ensuremath{{\mathrm{\,Me\kern -0.1em V\!/}c^2}}\xspace}
\newcommand{\gev}{\ensuremath{\mathrm{\,Ge\kern -0.1em V}}\xspace}
\newcommand{\gevc}{\ensuremath{{\mathrm{\,Ge\kern -0.1em V\!/}c}}\xspace}
\newcommand{\gevcc}{\ensuremath{{\mathrm{\,Ge\kern -0.1em V\!/}c^2}}\xspace}
\newcommand{\gevgevcc}{\ensuremath{{\mathrm{\,Ge\kern -0.1em V^2\!/}c^2}}\xspace}
\newcommand{\gevgevcccc}{\ensuremath{{\mathrm{\,Ge\kern -0.1em V^2\!/}c^4}}\xspace}

\begin{document}
\maketitle

\section{Introduction}
Flavour-Changing Neutral-Current (FCNC) decays of b-hadrons are forbidden 
at tree level in the Standard Model (SM) and are thus very suppressed.  As such, they are 
very sensitive to potential new particles entering the loops and affecting properties of the decays 
such as branching fractions and angular distributions. Consequently, the measurement of these processes allows to probe higher energy scales than direct searches. 

In recent years, several deviations with respect to SM predictions have been observed in this type of processes. 
Differential branching fraction measurements in \btosmm decays exhibit a trend towards values lower than the SM predictions in the di-lepton mass squared, $q^2$, 
region below the charmonium threshold~\cite{anomalies},
although theoretical predictions for these observables are affected by large hadronic uncertainties. 
Deviations have been also observed in the theoretically cleaner angular observables of the 
\BdToKstmm decay~\cite{LHCb-PAPER-2020-002}. 
Measurements of Lepton Flavour Universality (LU) in the decays \BuToKll and \BdToKstll, commonly referred to as \RK and \RKst, performed by LHCb exploiting the dataset collected during Run 1 of the LHC, also showed intriguing deviations with respect to the SM predictions~\cite{LUR1}: 
$\RK = 0.745^{+0.090}_{-0.074} \pm 0.036$  for $1.0<q^2<6.0 \gevgevcccc$,
$\RKst = 0.66^{+0.11}_{-0.07} \pm 0.03 $ for $0.045<q^2<1.1 \gevgevcccc$,
$\RKst = 0.69^{+0.11}_{-0.07} \pm 0.05 $ for $1.1<q^2<6.0 \gevgevcccc$,
where the first uncertainty is statistical and the second systematic. 
These results are compatible with the SM at the level of 2.6, 2.1 -- 2.3 and 2.4 -- 2.5 standard deviations, respectively. 
Interestingly, all the values were found to be lower than the SM universal prediction, 
and form a pattern of coherent deviations with the branching ratio and angular observables. 

In view of these anomalies in rare b-hadron decays, it is crucial to update the previous measurements with more data
and study new complementary modes. 
The most recent results from the LHCb experiment~\cite{LHCb-DP-2008-001} on LU tests in electroweak penguin decays 
are reported in the following.

\section{Lepton Flavour Universality tests at LHCb}
Leptons of different flavour couple identically to the electroweak bosons of the SM. 
This property has been tested with precision in transitions involving the first and second generation of quarks, e.g. in decays of $\phi$, \jpsi and \psitwos mesons~\cite{PDG}.
At LHCb, this property is studied in b-hadron decays and \bsll transitions with muons and electrons in the final state are of particular interest given the tensions observed in the differential branching ratio and angular measurements of these processes.
The ratio of branching fractions ($R_H$) for an exclusive $B \to H \ellp\ellm$ decay, with $\ell=\mu, e$, in bins of the di-lepton mass squared, $q^2$, is defined as
\begin{equation}
	R_H = \frac{\int \frac{d\Gamma(B \to H \mup\mun)}{dq^2}dq^2}{\int \frac{d\Gamma(B \to H e^+e^-)}{dq^2}dq^2},
\end{equation}
where $B$ and $H$ are a b- and s-hadron, respectively.
Theoretical uncertainties arising from the hadronic part of the decay cancel in the ratio, providing a very precise prediction.
In the SM, these ratios are computed to be unity with a per mile uncertainty~\cite{Fuentes}. 
Consequently these observables provide a null test of the SM and any deviation from unity would be a hint for New Physics (NP).

Experimentally, $R_H$ is extracted from the ratio of yields for both modes, corrected by the reconstruction and selection efficiencies.
Experimental effects differ significantly for electrons and muons at LHCb, and can induce large systematic uncertainties on the determination of absolute efficiencies. 
On one hand, the transverse momentum thresholds applied in the hardware trigger are much higher for electrons (e.g. 2400\mev in 2016) 
than for muons (e.g. 1800\mev in 2016)~\cite{trigger} to cope with the higher occupancy of the electromagnetic calorimeter (ECAL) compared to the muon chambers. 
This is mitigated by studying also events that are triggered by the hadron in the decay or by other particles in the event, so analyses are performed in various trigger categories. 
On the other hand, at the characteristic energies of b-hadron decays in the high-energy LHC environment, electrons emit a significant amount of Bremstrahlung radiation as they traverse the detector material, while this is not the case for muons. 
If a large amount of radiation is emitted before the LHCb magnet, the measurement of the electron momentum, which is based on the curvature of the particle in the magnetic field, does not account for the total momentum at production but is affected by the lost energy. 
A recovery procedure is in place to search for clusters in the ECAL that match the projection of the track direction before the magnet. The combination of the energy of the matching clusters with the momentum measured by the tracking system allows to recover the total electron energy. 
The performance of this recovery step is however limited by the resolution and occupancy of the ECAL, leading to a degraded momentum resolution for electrons as compared to muons. 

To reduce the impact of the experimental effects, $R_H$ is extracted from a double ratio, exploiting the resonant transition $B \to H \jpsi (\to \ellp\ellm)$
\begin{equation}
	R_{H} = \frac{\frac{N(B \to H \mu^+ \mu^-)}{N(B \to H J/\psi(\mu^+ \mu^-))}}
	{\frac{N(B \to H e^+ e^-)}{N(B \to H J/\psi(e^+ e^-))}}
	\times  \frac{\frac{\epsilon(B \to H e^+ e^-)}{\epsilon(B \to H J/\psi(e^+ e^-))}}
	{\frac{\epsilon(B \to H \mu^+ \mu^-)}{\epsilon(B \to H J/\psi(\mu^+ \mu^-))}
	}.
\end{equation}
The leptonic decay of the \jpsi is measured to be lepton universal with high precision~\cite{PDG} such that it cancels out in the ratio.
Using this method, reconstruction and selection efficiencies for the non-resonant modes need to be determined relative to the ones for the resonant mode, which shares the same final state, achieving a large cancellation of experimental uncertainties.
Moreover, the single ratio of branching fractions
\begin{equation}
	r_{J/\psi} = \frac{BR(B \to H J/\psi(\mu^+ \mu^-))}	{BR(B \to H J/\psi(e^+ e^-))},
\end{equation}
which is known to be unity with high precision, provides a cross-check of the control of the absolute efficiencies.

\section{Recent results}

An updated measurement of \RK has been recently performed at LHCb by reanalysing the Run 1 data and including part of the Run 2 sample, 
effectively doubling the number of studied b-hadron decays~\cite{RKR2}.
The analysis is perform in the region $1.1<q^2<6.0 \gevgevcccc$ in three trigger categories. 
A boosted decision tree trained on kinematic information is exploited to fight random combinations of tracks mocking the signal decay. 
Moreover, contamination from physics backgrounds arising from the misidentification of final state particles is suppressed by particle identification and invariant mass requirements. As an example, semileptonic b-hadron decays are efficiently rejected by requiring that the invariant mass of the $\Kp e^-$ pair is above the world-average \Dz mass.

\begin{figure}
	\centering
	\includegraphics[width=0.32\textheight]{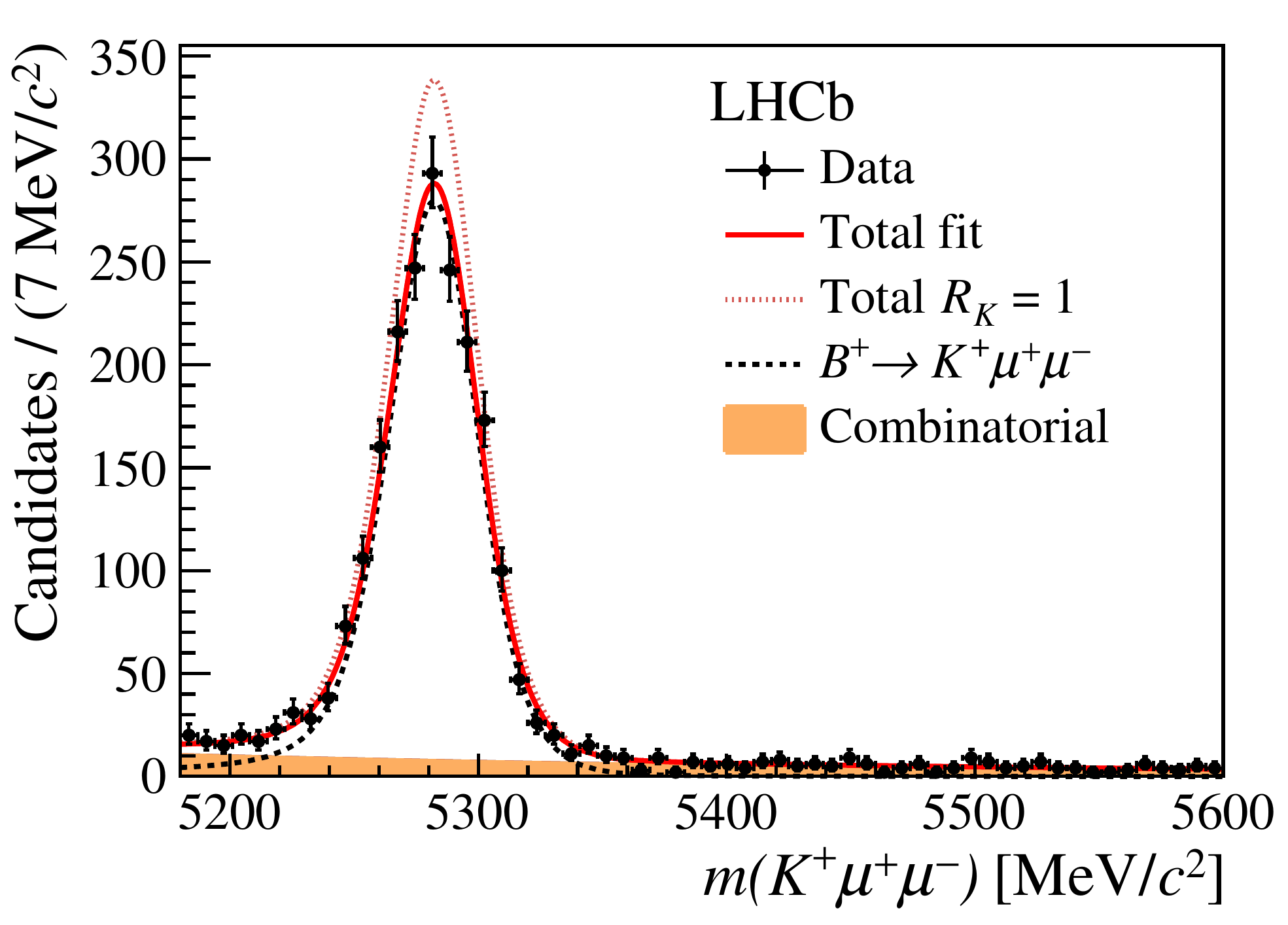}
	\includegraphics[width=0.32\textheight]{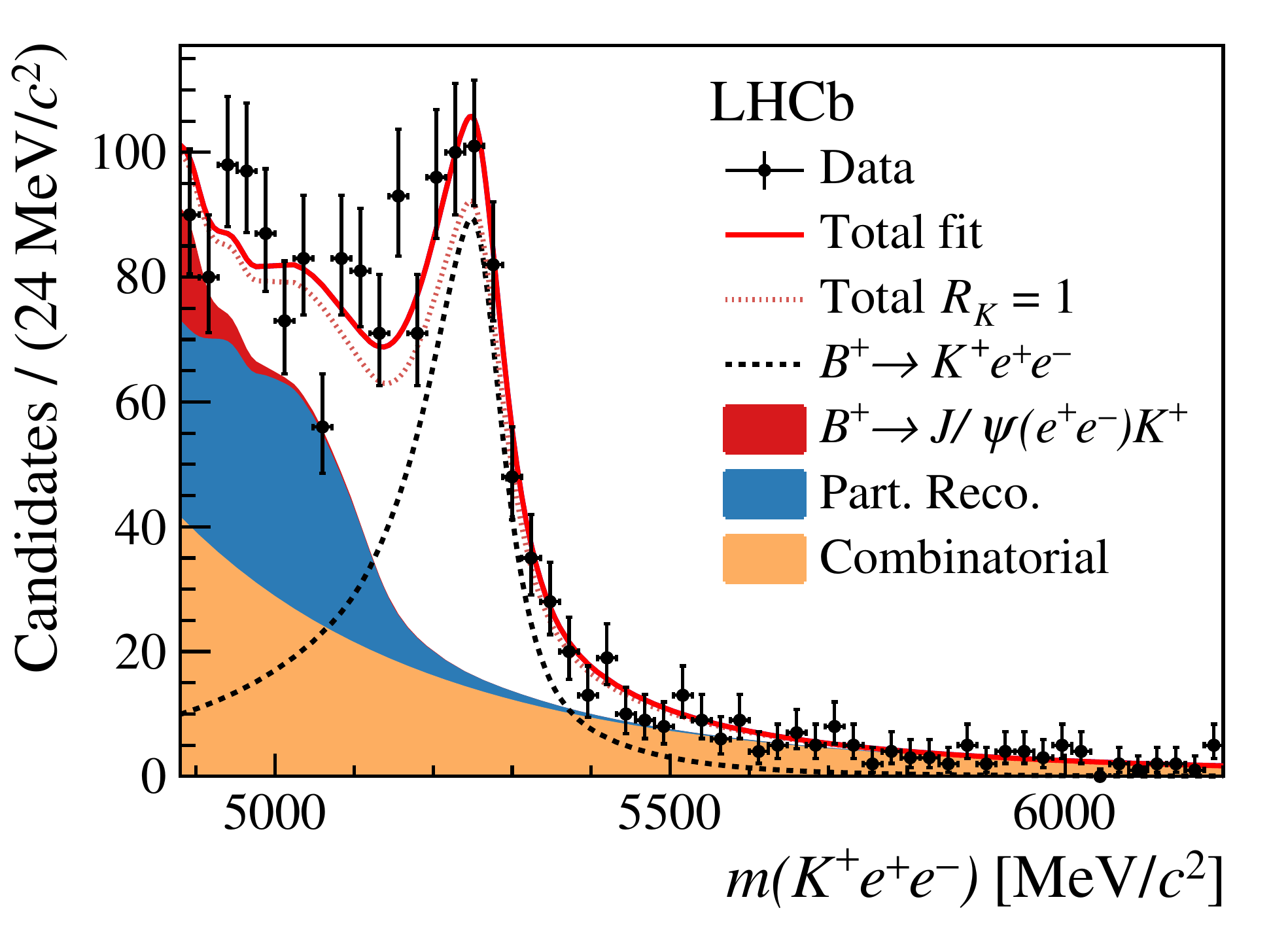}
	\caption{Invariant mass distribution of \BuToKpmm (left) and \BuToKpee (right) candidates. 
		The data is represented by black markers and the result of the fit by a solid red line. Different components are detailed in the legend.}
	\label{fig:RK_fits}
\end{figure}

A simultaneous fit to the invariant mass of \BuToKpmm and \BuToKpee candidates, shown in Fig.~\ref{fig:RK_fits}, is employed to extract \RK.
For the muon mode, a very clean signal of $1943 \pm 49$ is observed, with the only source of background arising from random combinations of tracks. 
The degraded mass resolution in the electron channel requires the extension of the fit range to a wider window, where more sources of background are present. Decays of the type $B \to \Kp e^+e^- X$, where X is not reconstructed, are generically referred to as partially reconstructed background, with \BdToKstee as the dominant source of these events. 
In addition, \BuToKJpsiee decays with a significant loss of energy can leak into the signal $q^2$ region and contaminate the low mass region.
Overall a good description is obtained and \RK is measured to be
\begin{equation}
	\RK = 0.846 ^{+0.060+0.016}_{-0.054-0.014},
\end{equation}
where the first uncertainty is statistical and the second systematic. 
This result is compatible with the one obtained in the previous analysis, confirming the tension with the SM prediction, which is found to be at the level of $2.5\sigma$.

The deviations of the measured \RK and \RKst values from the SM predictions motivate the measurement of LU in other decay modes. 
Of special interest are decays of b-baryons, which suffer from different experimental uncertainties and are sensitive to distinct
sources of NP due to the half-integer spin of the initial and final state particles. 

\begin{figure}
	\centering
	\includegraphics[width=0.32\textheight]{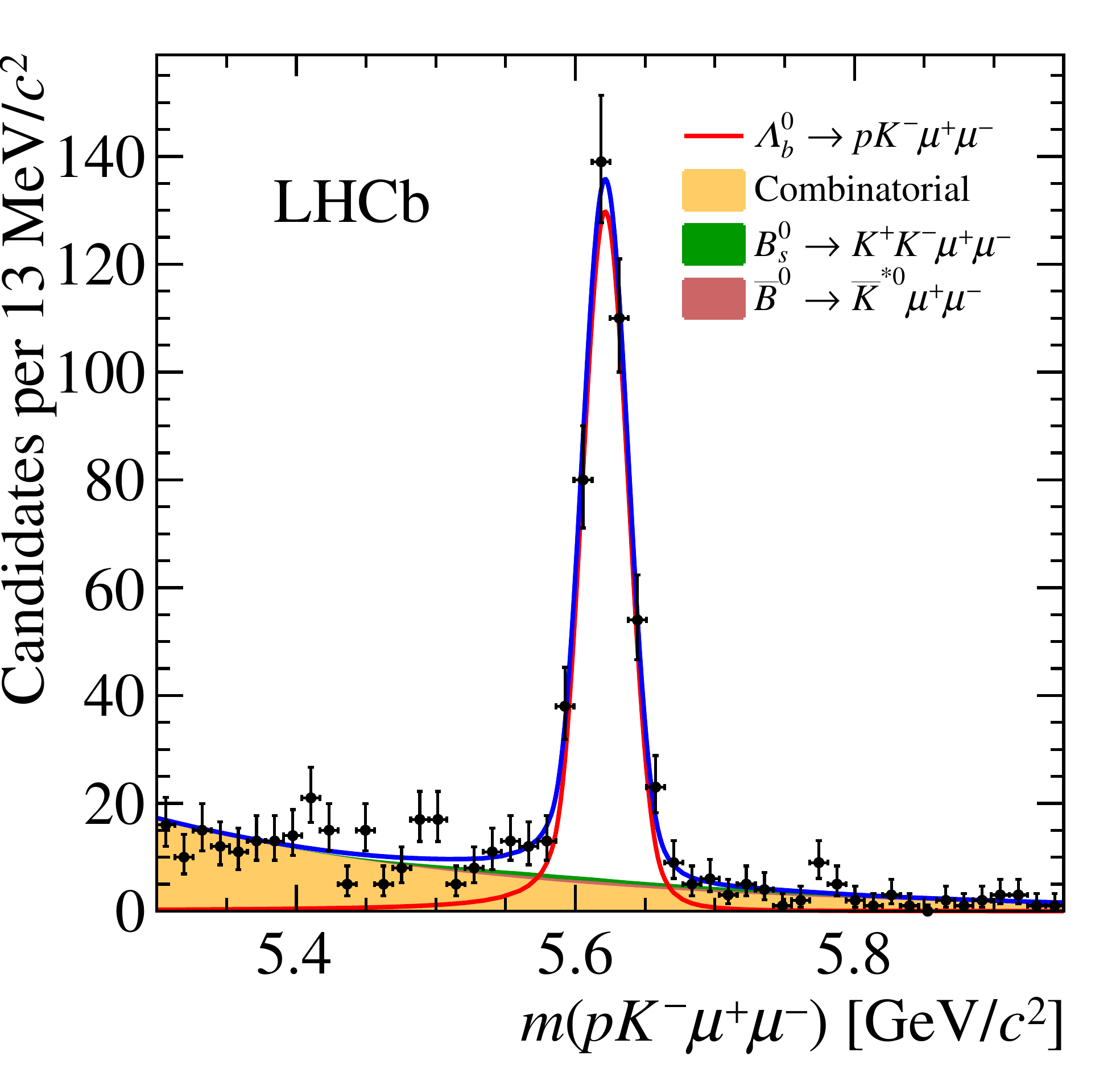}
	\includegraphics[width=0.32\textheight]{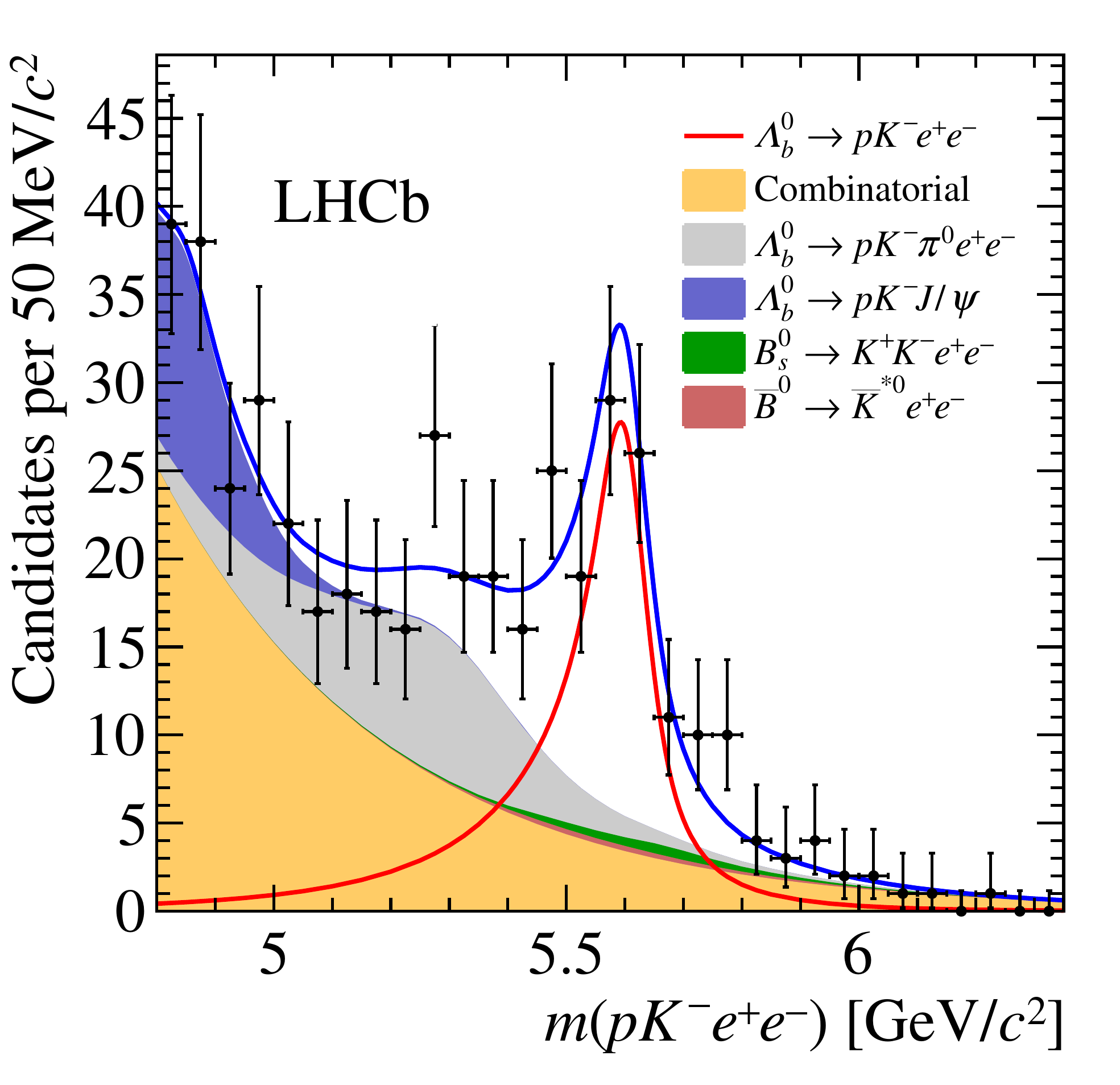}
	\caption{Invariant mass distribution of \LbTopKmm (left) and \LbTopKee (right) candidates. 
		The data is represented by black markers and the result of the fit by a solid blue line. Different components are detailed in the legend.}
	\label{fig:RpK_fits}
\end{figure}

Very recently, LHCb has tested LU in the transition \LbTopKll for the first time (\RpK), exploiting the data collected during Run 1 and 2016~\cite{RpK}. 
The selection of signal candidates follows the strategy of the \RK analysis. In this case though, only two trigger categories are employed.
The measurement is performed in the region $0.1 < q^2 < 6.0 \gevgevcccc$ and $m(p\Km) < 2600 \mev$.
The quantity $m(p\Km)$ refers to the invariant mass of the $p\Km$ pair, where a number of states are observed to contribute.
The single ratio of branching fractions for the resonant mode is found to be compatible with unity, $\invrJpsi = 0.96 \pm 0.05$, with the uncertainty dominated by systematic effects. 
This quantity is studied as a function of various kinematic variables. 
All the trends are observed to be flat and consistent with unity, confirming that the determination of the efficiency is under control in different kinematic regions, as needed to measure \RpK.
The latter is extracted from a simultaneous invariant mass fit to \LbTopKmm and \LbTopKee candidates, shown in Fig.~\ref{fig:RpK_fits}. 
Similarly to the \RK case, the muon mode is very clean and is only polluted by combinatorial background, while the electron mode is affected by more backgrounds. In this case, partially reconstructed background arises mostly from $\Lb \to p\Km\piz \epem$ decays, where the \piz is not reconstructed. Moreover, the b-meson decays $\Bs\to\Kp\Km\epem$ and $\overline{\B}^0 \to\overline{K}^{*0}\epem$ are additional sources of contamination due to hadron misidentification. 
In total, $444 \pm 23$ \LbTopKmm and $122 \pm 17$ \LbTopKee events are observed. 
The value of \RpK in the region $0.1 < q^2 < 6.0 \gevgevcccc$ and $m(p\Km) < 2600 \mev$ is found to be
\begin{equation}
	\RpK = 0.86^{+0.14}_{-0.11} \pm 0.05,
\end{equation}
where the first uncertainty is statistical and the second systematic. 
This measurement is compatible with the SM universal prediction and with the previous measurements of \RK and \RKst, 
which lay below the SM. More data is needed to confirm or dismiss the presence of LU-violating NP in \bsll transitions.

\section{Conclusions and prospects}
Decays involving \bsll transitions are an excellent laboratory to search for NP. 
A coherent pattern of deviations with respect to the SM predictions, involving branching fractions, angular observables and LU tests, is forming during the last years and could be a first hint for a new interaction of Nature~\cite{patterns-LQ}.
Tests of LU in \bsll decays provide the most precise theory predictions, enabling a null test of the SM. 
Updated results from LHCb on \RK and for the first time on \RpK are consistent with previous measurements, maintaining the observed tensions with the SM. However more data is needed to confirm the presence of NP. 
Interestingly, the deviations observed in \bsll decays can be explained by NP models that also address tensions in tree-level $\bquark\to\cquark\ell\nu$ transitions~\cite{patterns-LQ}.

LHCb has already collected a sample of b-hadrons twice as large as used in the measurements presented here, which will allow to reach a precision of 4, 5 and $10\%$ on \RK, \RKst and \RpK, respectively. 
The study of additional modes with this sample, such as \BsToPhill and \BuToPill, will provide independent cross-checks of the observed tensions.
Further in the future, the upgraded LHCb detector and the Belle II experiment will enable a clear observation of NP if the central values of LU observables stay the same~\cite{LHCb-U2-Belle2}.

\end{document}